\documentclass[apjl]{emulateapj}
\usepackage{amsmath}

\slugcomment{ApJ accepted}
\shorttitle{Off-Axis GRB and Orphan GRB afterglow}
\shortauthors{Urata et al.} 
\begin{document}
\title{Extremely Soft X-ray Flash as the indicator of off-axis orphan GRB afterglow}

\author{
Yuji~\textsc{Urata}\altaffilmark{1}, 
Kuiyun~\textsc{Huang}\altaffilmark{2},
Ryo~\textsc{Yamazaki}\altaffilmark{3},
Takanori~\textsc{Sakamoto}\altaffilmark{3}
}

\altaffiltext{1}{Institute of Astronomy, National Central University, Chung-Li 32054, Taiwan, urata@astro.ncu.edu.tw}
\altaffiltext{2}{Department of Mathematics and Science, National Taiwan Normal University, Lin-kou District, New Taipei City 24449, Taiwan} 
\altaffiltext{3}{Department of Physics and Mathematics, College of Science and Engineering, Aoyama Gakuin University, 5-10-1 Fuchinobe, Chuo-ku, Sagamihara-shi, Kanagawa 252-5258, Japan}
	
\begin{abstract}

  We verified the off-axis jet model of X-ray flashes (XRFs) and
  examined a discovery of off-axis orphan gamma-ray burst (GRBs)
  afterglows. The XRF sample was selected on the basis of the
  following three factors: (1) a constraint on the lower peak energy
  of the prompt spectrum $E^{src}_{obs}$, (2) redshift measurements,
  and (3) multi-color observations of an earlier (or brightening)
  phase. XRF020903 was the only sample selected basis of these
  criteria.  A complete optical multi-color afterglow light curve of
  XRF020903 obtained from archived data and photometric results in
  literature showed an achromatic brightening around 0.7 days. An
  off-axis jet model with a large observing angle (0.21 rad, which is
  twice the jet opening half-angle, $\theta_{jet}$) can naturally
  describe the achromatic brightening and the prompt X-ray spectral
  properties.
  This result indicates the existence of off-axis orphan GRB afterglow
  light curves. Events with a larger viewing angle
  ($>\sim2\theta_{jet}$) could be discovered using an 8-m class
  telescope with wide field imagers such as Subaru Hyper-Suprime-Cam
  and the Large Synoptic Survey Telescope.

\end{abstract}

\section{Introduction}

Long gamma-ray bursts (GRBs) are believed to occur when a very massive
star dies in a highly energetic supernova forming a black hole and
producing a relativistic jet.  Because of the release of a large
amount of isotropic equivalent energy (${\it E}_{\rm
  iso}\sim10^{52-55}$ erg) release in the short prompt gamma-ray phase
(typically ranging from several seconds to several tens of seconds),
the consideration of jet collimation of GRBs is necessary to explain
the radiation mechanism from compact sources (e.g., massive stars
and/or mergers). This necessity is supported by achromatic temporal
breaks (also known as jet breaks) in the afterglow light curves of the
GRBs.  Ultra-relativistic collimation and a jet structure are required
to explain the light curve temporal breaks
\citep[e.g.,][]{sari}. However, no direct observational evidence
exists for this jet collimation.

Off-axis orphan GRB afterglows are produced as a natural consequence
of GRB jet production \citep{rhoads}.  The production of these
afterglows is as follows: GRBs are collimated with rather narrow
opening angles, and the afterglow that follows can be observed over a
wider angular range. While the GRB and the early afterglow are
collimated to within the original jet opening angle, the afterglow in
the late phase can still be observed by an off-axis observer after the
jet break.  The Lorentz factor, $\Gamma$, is a rapidly decreasing as
function of time. This means that an observer at $\theta_{obs}$ cannot
see the prompt gamma-ray emissions when $\theta_{obs} > \theta_{jet}$
but can detect an afterglow once $\Gamma^{-1} $ equals
$\theta_{obs}$. Here, $\theta_{jet}$ is the jet opening half angle. As
the typical emission frequency and flux decrease with time (while the
jet opening half angle $\theta_{jet}$ increases with time), observers
at larger viewing angles will detect fainter afterglows at longer
wavelengths (e.g. optical and radio). Hence, the expected properties
of GRB orphan afterglows are as follows: (1) prompt emissions in the
high-energy band are absent, (2) their brightness is fainter than that
of on-axis GRB optical afterglows, (3) they have the same optical
color as on-axis afterglows, and (4) they show host galaxy properties
similar to those of on-axis GRBs. The afterglows are characterized by
three-component light curves with rising, peaking, and rapidly
decaying phases.  In this case, events intermediate between classical
hard GRBs and off-axis orphan GRB afterglows should exist.

A candidate for the intermediate events is X-ray flashes (XRFs) as
\citet{offaxis, yamazaki} described in their off-axis jet model for
explaining their nature.  XRFs were recognized by the Wide Field
Cameras \citep[WFC, 2-28 keV;][]{wfc} onboard the {\it BeppoSAX}
satellite \citep{heise}. The observed X-ray temporal and spectral
properties of XRFs in the prompt phase do not show any differences
relative to those of GRBs, except for the considerably lower energy
values of the peak of the $\nu F_{\nu}$ spectrum in the observer's
frame.  A large number of XRF samples were provided by {\it BeppoSAX}
and High Energy Transient Explorer 2 ({\it HETE-2}), both of which
employed wide field X-ray cameras $--$ the WFC onboard BeppoSAX and
the Wide-Field X-ray Monitor \citep[WXM, 2-25 keV][]{shirasaki}
onboard {\it HETE-2}.  The majority of {\it HETE-2} samples (nine out
of 16 XRFs) show a low energy of spectral peak energy
$E^{obs}_{peak}<20$ keV \citep{sakamoto05}.  The number of XRFs
detected by {\it HETE-2} was comparable and relatively larger than
that of GRBs indicating that XRFs represent a large portion of the
entire GRB population \citep{sakamoto05}.  The observational
properties of XRFs can be interpreted as being associated with the
same phenomenon as classical hard GRBs and as being representative of
the extension of the GRB population to low peak-energy events
\citep{kippen,sakamoto05}.
To explain the aforementioned prompt observational properties, three
models have been proposed for XRFs: a high redshift origin
\citep{heise03}; the off-axis jet model
\citep{offaxis,yamazaki,zhang04,lamb05}, which is equivalent to the
unification scenario of AGN galaxies; and intrinsic properties (e.g.,
a subenergetic or inefficient fireball), which may also produce
on-axis orphan afterglows \citep{huang02}.

The {\it Swift} satellite has also been detecting many X-ray rich GRBs
(XRRs) and XRF samples \citep{sakamoto08}. However, the XRF samples
tend to be at the high-end of the $E^{obs}_{peak}$ distribution of
{\it BeppoSAX} and {\it HETE-2}. This is because of the relatively
higher energy coverage of the Burst Alert Telescope \citep[BAT, 15-150
keV;][]{bat} onboard {\it Swift}.  Although the Monitoring of All-sky
X-ray Image \citep[MAXI;][]{maxi} attached to the International Space
Station has been detecting XRFs \citep{serino}, there has been no
appropriate follow-ups because of poor position determination.  Hence,
XRF studies have stagnated because of the lack of soft-X-ray
monitoring instruments and intensive multiwave length follow-up
observations.
 
In this paper, we investigated the characteristics of XRFs on the
basis of redshift measurements, multifrequency afterglow monitoring,
and spectral peak measurements $E^{src}_{peak}$. The main objective
was to verify the off-axis jet model and to provide feedback to
ongoing and planned optical untargeted time-domain surveys by using
Subaru Hyper-Suprime-Cam \citep{hsc} and the Large Synoptic Survey
Telescope (LSST), which have considerable potential for detecting
off-axis orphan GRB afterglows.

\section{Samples of X-Ray Flash}

\begin{figure}
\epsscale{1.0}
\plotone{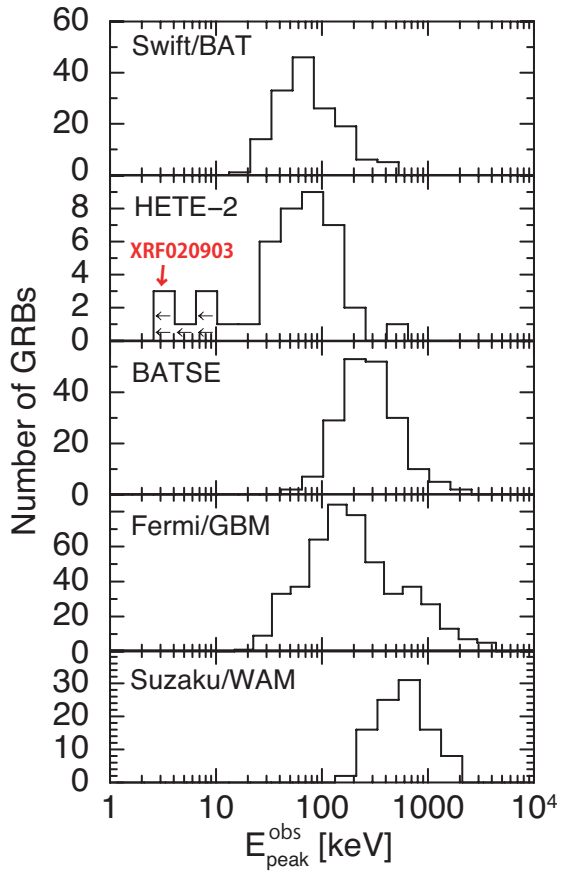}
\caption{Distribution of $E^{obs}_{peak}$ for the {\it Swift}/BAT, {\it HETE-2}, the BATSE, the {\it Fermi}/GBM, and the {\it Suzaku}/WAM samples. XRF020903 showed the lowest $E^{obs}_{peak}$ among all five samples.}
\label{fig:ephist}
\end{figure}

We considered possible XRFs for our study and quickly realized XRF
020903 was the only event that has (1) either a measurement or an
upper bound on the peak energy of the prompt spectrum
$E^{src}_{peak}$, (2) a measured redshift, and (3) multicolor
afterglow observations adequately cover the early afterglow phase when
achromatic brightening of the afterglow might occur.
This was one of the XRFs detected by {\it HETE-2} and the first events
for which an optical afterglow was detected and a spectroscopic
redshift ($z=0.251$) was determined \citep{radio}.  The prompt
emission had the lowest intrinsic spectral peak energy
$E^{src}_{peak}$ of $3.3^{+1.8}_{-1.0}$ keV among all the XRF samples.
Here, we employed the value estimated by the constrained Band function
with the 90\% confident level \citep{020903}. In accordance with the
report of \citet{020903}, the light curve in the prompt phase
exhibited a double-peak structure and a lack of signals above 10
keV. Figure \ref{fig:ephist} shows histograms of the spectral peak
energy in the observer frames $E^{obs}_{peak}$ detected by {\it
  Swift}/BAT \citep{lien15}, {\it HETE-2} \citep{sakamoto05}, the
Burst and Transient Source Experiment \citep[BATSE;][]{kaneko06}, the
{\it Fermi} Gamma-ray Burst Monitor \citep[{\it
    Fermi}/GBM;][]{kienlin14}, and {\it Suzaku} Wide-Band All-Sky
Monitor (WAM). The WAM $E^{obs}_{peak}$ distribution is produced using
values, which are available in GCN circulars up to December
2014\footnote{http://www.astro.isas.jaxa.jp/suzaku/HXD-WAM/WAM-GRB/results/gcn.html}.
The standard analysis procedure of WAM data \citep{wam} are described
in literature \citep[e.g.,][]{ohno08,100414a}. For comparison with a
large volume of data, we used $E^{obs}_{peak}$ instead of intrinsic
spectral peak energies.  The lowest intrinsic spectral peak energy
determined from the relation $E^{src}_{peak}$=(1+$z$)$E^{obs}_{peak}$
of XRF020903 stands out from all GRB populations. The X-ray to the
$\gamma$-ray fluence ratio of 5.6 qualified this burst as an
XRF. Here, we used the XRF definition by \citet{sakamoto05} for {\it
  HETE-2} events.

Effective follow-up observations in the optical band were made using
wider-field-of-view (FOV) instruments reported by \citet{radio} and
\citet{ctio}. Because of a delay in the position alert and a large
position error, the light curve sampling was sparse around the
possible rebrightening epoch reported by \citet{ctio}. To describe the
rebrightening epoch by using multicolor data, we added Subaru archive data,
as described in \S 3. 

\section{Observations and Data collection}

We collected data for the XRF~020903 afterglow by using Subaru archive
data and photometric results obtained from literature
\citep{ctio,radio}.  To perform accurate optical photometry by
removing contamination from the host galaxy, we also employed
Panoramic Survey Telescope and Rapid Response System 1 (Pan-STARRS1)
data. The individual data collections are summarized in the following
subsections.

\begin{figure*}
\plotone{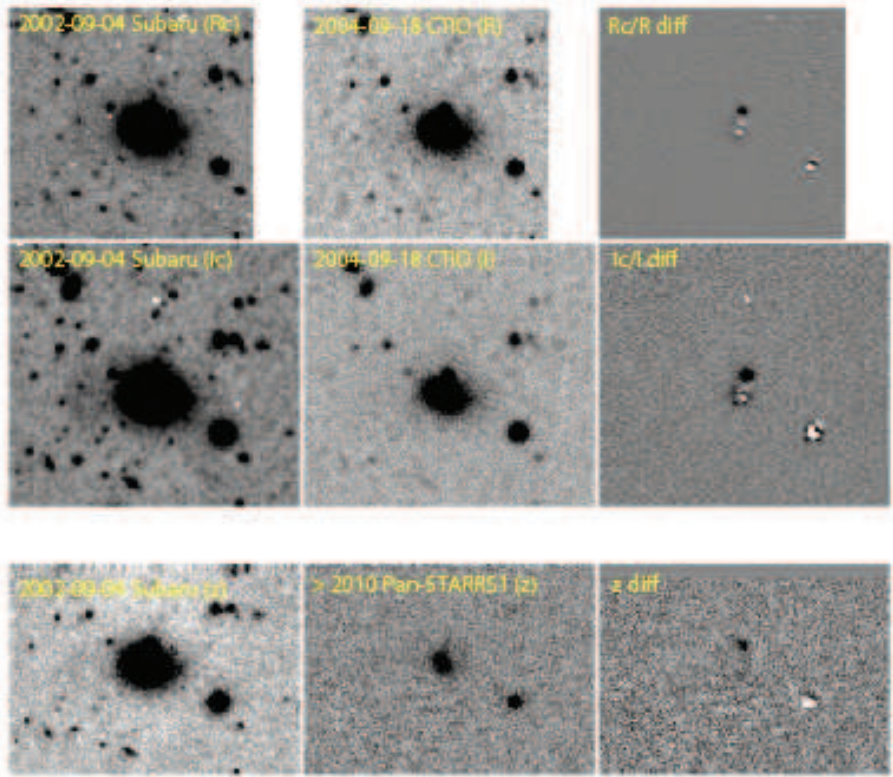}
\caption{$Rc/R$, $Ic/I$, and $z$-band images of the XRF020903 field. The first column shows the Subaru images taken in the $Rc$ (top), $Ic$ (middle), and $z$-band (bottom). Reference images for removing the host galaxy component by using the algorithm of \citet{alard} are shown in the second column. The top-center, middle-center, and bottom-center panels show the reference images taken with the CTIO $R$-band, CTIO $I$-band and PS1 $z$-band filters. The third column shows $Rc/R$, $Ic/I$, and $z$-band differential images, respectively. 
  \label{fig:image}}
\end{figure*}

\subsection{Subaru Suprime-Cam}

The entire {\it HETE-2} position error region was imaged by
Suprime-Cam attached to the 8.2-m Subaru telescope. The Suprime-Cam
camera consists of ten high-sensitivity 2k$\times$4k CCDs and covers a
field of view of $34'\times27'$ \citep{sc}. The first epoch of
observation on 2002 September 3 (0.16 day after the burst) involved an
$Rc$-band filter. Although the observation was performed under
marginal airmass conditions (from 2.55 to 2.75), the wide-field imager
with the 8-m-class telescope provided the deep $Rc$-band images with
three sets of 180 s exposure.
Subsequently, three color observations were made on the night of 2002
September 4 by using $Rc-$, $Ic-$, and $z'$-band filters. During the
night observation, two epochs of $Ic-$ and $z'-$ bands imaging and
three epochs of $Rc$-band monitoring were also conducted to check the
short-term variability of the afterglow. The $Ic-$ and $z'-$ band
observations were made under the reasonable observing condition (e.g.,
in the airmass range from 1.34 to 1.97). The second epoch of $Rc$-band
imaging was also performed under the reasonable observing condition
(e.g., for an airmass of 1.32). By contrast, the first and third
epochs of $Rc$-band observations were made under the marginal airmass
conditions ($2.65-2.22$ for the first epoch and $1.97-2.27$ for the
third epoch). These observations were performed with using appropriate
dithering techniques to fill up the chip gaps in the camera.

The Subaru-XMM Deep Survey (SXDS) team attempted to obtain deeper
reference images with $Rc$-band filter on the night of October 9 and
10. However, the results were marginal because of poor weather
conditions and only two set of images taken with 180 s exposure on
2002 October 9 were available for scientific analysis. Table \ref{obs}
shows the log of observations for the available images. Because of the
limited data set, searching for a counterpart by using only the
Suprime-Cam data was difficult. All raw images and related calibration
data are available on the SMOKA \citep[Subaru Mitaka Okayama Kiso
Archive system][]{smoka}.

\subsection{CTIO}

We obtained $R$ and $I$ images taken by the wide field MOSAIC II
camera on the Cerro Tololo Inter-American Observatory (CTIO) Blanco
4-m telescope from the NOAO archive system. These data were taken on
2004 September 18 with 1800 s exposure (360s$\times$5) for the
$R$-band, and on 2004 September 9 with 1800 s exposure (450s$\times$4)
for the $I$-band, when was sufficiently late to estimate the host
galaxy contribution on the Subaru images. The data were processed
using the pipeline of the NOAO archive system
\citep{noaopipe1,noaopipe2}. These two stacked images were also used
to remove the host galaxy contamination for describing the late phase
afterglow temporal evolution presented by \citet{ctio}.  We also
attempted to obtain $B$, $R$ and $I$ images taken by \citet{ctio} on
2002 September 4 and 9 ($\sim$0.66 day and 5.7 day after the burst,
respectively). However, these imaging data were unavailable because of
a limitation of the tape reader device on the NOAO archive system.

\begin{figure}
\plotone{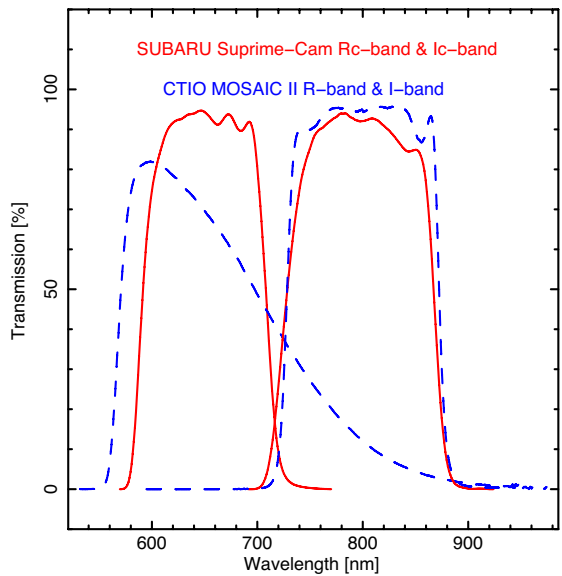}
\caption{$R$-band and $I$-band transmission curves for Subaru Suprime-Cam (red solid line) and CTIO MOSAIC II (blue dashed line). The data were obtained from their respective instrumental Web-pages.
\label{fig:filter}}
\end{figure}

\subsection{Pan-STARRS1 $z$-band image}

Pan-STARRS1 (PS1) $z'$-band images were also obtained to determine the
host galaxy contamination on the Subaru $z$-band images taken in 2002
September 4. The PS1 telescope has a 1.8-m diameter primary mirror,
and it is located at the summit of Mt. Haleakala on Maui. The site and
optics deliver a point-spread function (PSF) with a full-width at
half-maximum of about 1 arcsec, over a seven square degree field of
view. The PS1 was used to conduct a $3\pi$ survey of the entire sky
north of $-30^{\circ}$ in $g'$-, $r'$-, $i'$-, $z'$- and $y$-band
\citep{ps1a,ps1b,ps1c}. Because of the PS1 3$\pi$ surveys strategy,
the XRF020903 field was covered during the 3.5 years of survey
starting from 2010.  The images were processed by the Image Processing
Pipeline \citep{ipp}, and a deeper stacked $z$-band image was
generated by the SWarp software \citep{swarp}.
The PS1 3$\pi$ catalogs were also used to perform photometric
calibration for the $z$-band images taken by Subaru.

\subsection{Late phase optical observations}

Late optical afterglow in the $R$-band was also monitored by the 1.82
m Copernicus telescope at Mount Ekar (Asiago, Italy) on 2002 September
29 (26.5 days), and by the 3.5m Telescopio Nazionale Galileo (TNG) at
La Palma with $V$-, $R$-, and $I$-band filters on 2002 October 2 (29.5
days). Additional $BVRI$ images were obtained with the Danish 1.5-m
telescope at the La Silla Observatory between 2002 October 10 (36.8
days) and 14 (40.7 days) and then on October 26 (52.6 days).
Differential images without host galaxy contamination were produced
from late time images taken in 2004 through image convolution and by
using subtraction methods of \citet{alard}.  These photometric results
for the optical afterglow against with the secondary standard stars
from the list of \citet{standard} were reported by \citet{ctio}.

\subsection{Radio observations}
Very Large Array (VLA) observations were performed at 8.5 GHz on 2002
September 27.22 UT and radio afterglow was detected. Further
monitoring observations were made with the VLA over 370 days at
frequencies of 1.5, 4.9, 8.5, and 22.5 GHz \citep{radio}.
The Very Long Baseline Array also observed the radio afterglow and
determined the position accurately, as $\alpha_{2000}=22^{\rm
  h}48^{\rm m}42^{\rm s}.33912\pm0^{\rm s}.00003$,
$\delta_{2000}=-20^{\circ}46'08".945\pm0".0005$ \citep{radio}.

\section{Data Analysis and Results}

\subsection{Reduction}

The basic reduction of the Suprime-Cam data was performed using the SDFRED
package \citep{sdfred}.  This entailed bias subtraction and flat-fielding in the 
$Rc$-, $Ic$- and $z'$-band by using a sky flat constructed from the median of
the dithered science frames. After performing distortion correction for each
object frame, we stacked the frames for each epoch and for each band path filter by
considering the median.  We performed astrometric calibration for the stacked
images against the 3$\pi$ catalog of PS1.

For other data, we used the reduced data as previously summarized,
except photometric calibration.  The absolute photometric calibration
for $Rc$ and $Ic$ was performed using standard stars from the list of
\citet{standard}.  For the photometric calibration of $z$-band images,
we used the PS1 $3\pi$ catalog and selected unsaturated stars around
the afterglow location.

\begin{deluxetable}{rcc}
\tablewidth{0pt}
\tabletypesize{\scriptsize} 
\tablecaption{Log of Subaru follow-up observations.\label{obs}}
\tablehead{
\colhead{Delay (Days)}&
\colhead{Filter}&
\colhead{Flux density ($\mu$Jy)}}
\startdata
 0.1645 & $Rc$ & $ 21.22 \pm 1.09 $ \\
 0.8609 & $Rc$ & $ 34.87 \pm 1.38 $ \\
 1.0087 & $Rc$ & $ 35.00 \pm 1.39 $ \\
 1.1378 & $Rc$ & $ 29.57 \pm 1.28 $ \\
35.78945 & $Rc$ & $3.96\pm0.46$ \\ \hline
 0.8991 & $Ic$ & $ 49.21 \pm 1.82 $ \\
 1.0436 & $Ic$ & $ 41.08 \pm 1.67 $ \\ \hline
 0.9569 & $z$  & $ 55.67 \pm 2.06 $ \\
 1.0955 & $z$  & $ 54.10 \pm 1.96 $ 
\enddata                               
\end{deluxetable}

\subsection{Afterglow light curves}

\begin{figure*}
\plotone{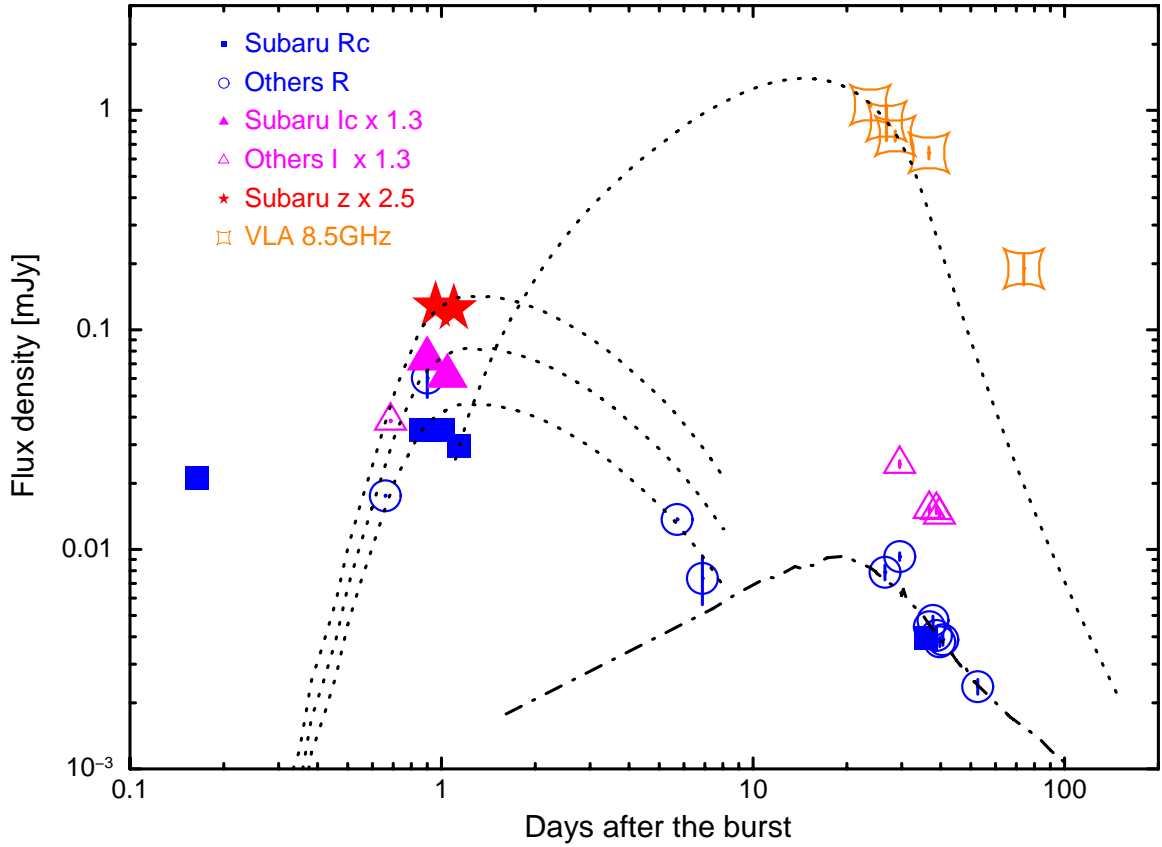}
\caption{$Rc/R$, $Ic/I$, $z'$-band and VLA 8.5GHz light curves of XRF020903. The dotted lines show the off-axis jet model functions described in \S5.1. The dash-dotted line indicates the SN component reported by \citet{ctio}.
\label{fig:lc}}
\end{figure*}

As \citet{ctio} reported, the host galaxy of XRF020903 is dominated
and contaminated in afterglow measurements. Hence, to remove the
galaxy contamination, we generated differential images using the
HOTPANTS software, which employs an algorithm presented by in
\citet{alard}.  We also used special-purpose software based on the
same algorithm and tuned to the Subaru/Suprime-Cam data
\citep{sdftran} to check the consistency.  For $Rc$- and $Ic$-band
data of Subaru/Suprime-Cam, we used the $R$- and $I$-band images taken
by the CTIO in 2004. The depth of these images was comparable to that
of Suprime-Cam images. For the $z$-band, the reference image generated
using the PS1 3$\pi$ data was shallower than that of the Subaru
image. However, there were the significant signals at the GRB
afterglow position in the generated image (Figure
\ref{fig:image}). Hence, we also generated the differential images by
using the same code.  As shown in Figure \ref{fig:image}, the quality
of the differential images is appropriate for estimating the
brightness of an afterglow component by the standard aperture
photometry.  There are appreciable differences between the
transmission curves of the Suprime-Cam $Rc$ and CTIO $R$ bands.
Figure \ref{fig:filter} shows the transmission curves for the
Suprime-Cam ($Rc$ and $Ic$) and CTIO ($R$ and $I$) filters. This transmission difference
causes a systematic difference of 17\% in the magnitudes for an object
with a power-law spectrum of index $-1$.

Figure \ref{fig:lc} shows the optical light curves for our Subaru
data and photometric results reported by \citet{ctio}; the radio
afterglow light curve at 8.5GHz \citet{radio} is also shown.  There is a
significant signal at 0.165 days after the burst. The use of only one epoch and
single-color observations make it difficult to describe the afterglow temporal evolution.
When we consider a simple direct connection to the first epoch of
$R$ band photometry (0.660 days) made by CTIO, the $Rc/R$ band light
curve is flat.  The $Rc/R$- and $Ic/I$- band light curves
show a clear rapid rebrightening between 0.7 and 0.9 days. The
equivalent rising power law index $\alpha$ ($F_{\nu} \propto
t^{\alpha}$) are $\sim2.6$ in the $Rc/R$ and $\sim2.3$ in the $Ic/I$ band.
The multiepochs observations of the Subaru on 2002 September 4 show
a flatter and decaying light curve for all three bands. The $Rc/R$ band light curve
shows continuous decay up to $\sim$7 days. This indicates that the
rebrightening peak was around 0.8$\sim$0.9 days after the burst.
As \citet{ctio} demonstrated, there is another late-phase
rebrightening peaking at $\sim20$ days in the $R$ band light curve,
which was interpreted as the associated supernova component. 
The late phase $Rc$-band image taken by Subaru on 2002 October 9 also
detected this supernova component in the differential image, and this
component is consistent with that observed in the $R$-band within the
systematic error.

\subsection{Spectral flux distributions}

\begin{figure}
\plotone{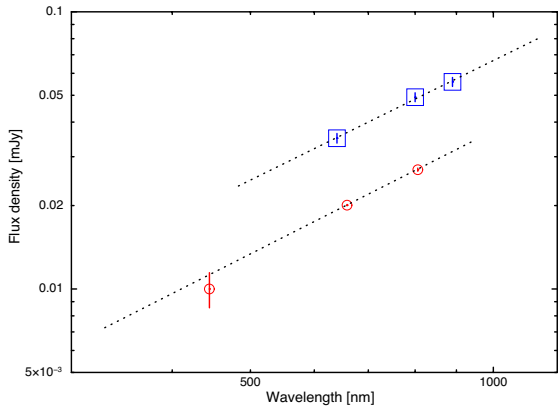}
\caption{Spectral flux distribution of the XRF020903 afterglow at 0.695 (red circle point) and 0.899 (blue square points) day. The dashed lines show the best-fit model functions.
\label{fig:sed}}
\end{figure}

The multiband observations of XRF 020903 were used to determine the
spectral flux distribution (SED) at 0.7 and 0.9 day after the
burst.
To remove the effects of the Galactic interstellar extinction, we used
the reddening map of \citet{schlafly}. Because not all multiband
observations were performed exactly at the same epoch, their
corresponding fluxes were rescaled to assume a power-law function
($F_{\nu} \propto t^{-\alpha}$). In particular, the brightening phase
around 0.7 day was substantially affected.
To determine the spectral energy distribution (SED) at 0.7 day, we
fixed the time at the epoch of the $B$-band observation (0.695
days). Because the $Rc/R$- and $Ic/I$-band observations were made
earlier than $B$ band observations, and the brightening index for
$Rc/R$ and $Ic/I$ were estimated as previously described (\S 4.2).
In figure \ref{fig:sed}, we plot the SED obtained on the basis of the
photometric results for $BRI$ bands provided by the CTIO observations.
The SED is well fitted by the power-law function as $f(\nu)\propto
\nu^{-\beta}$, where $f(\nu)$ is the flux density at frequency $\nu$,
and $\beta$ is the spectral index.  We have obtained the $\beta$ at
0.695 day as $1.48\pm0.06$.
Similarly, we generated the SED at 0.899 day on basis of the on
$Rc$-,$Ic$-, and $z'$-band results and obtained a $\beta$ value of
$1.43\pm0.08$, which is consistent with the value for 0.695 day.
These results imply that the rebrightening is achromatic.

\section{Discussion}

On the basis of the results for XRF020903 described in the preceding
section, we discuss the off-axis jet model in the following sections
because this model provides a reasonable explanation for this
particularly soft XRF. Furthermore this model can explain both the
extremely soft prompt emission features and achromatic light curve
brightening altogether.

\subsection{Off-axis modeling of afterglow light curves}

To perform light curves and SEDs modeling for XRF020903, we employed
the boxfit code \citep{boxfit} that involves two-dimensional
relativistic hydrodynamical jet simulations for determining the burst
explosion parameters, including the off-axis viewing angle and the
synchrotron radiation parameters for a homogeneous circumburst medium.

\begin{figure}
\plotone{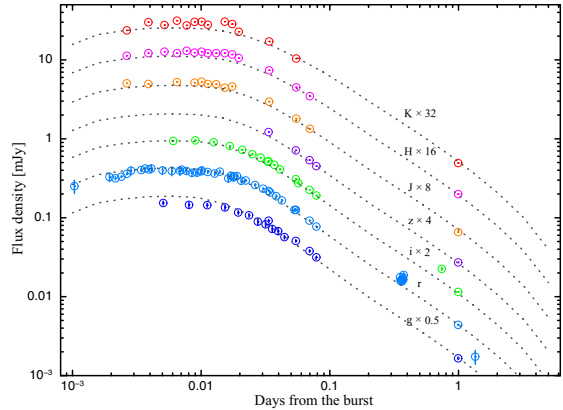}
\caption{XRF080330 multi-band light curves with the off-axis jet mode function (dotted lines). The optical and near-infrared data points reported by \citet{080330} were used for the light curve modeling.
\label{fig:080330}}
\end{figure}

In the modeling, we added 17\% of the systematic errors to the
$R$-band light curve because of the large transmission differences, as
shown in Figure \ref{fig:filter}.  To perform light curve modeling, we
focused on the data between 0.6 and 8 day because the observed optical
light curve showed achromatic brightening between 0.7 and 0.8 day
after the burst, which is unusual among the well-observed
afterglows. The later phase (from $\sim$20 days) was excluded in the
modeling because it is likely to be the SN component \citet{ctio}. We
also excluded the single data point at 0.16 day because the sparse
monitoring and single detection at 0.16 day make it difficult to
decode the light curve between 0.16 and 0.7 day. The consideration of
different components (one component faded away before 0.6 day and the
other was rebrightening) was reasonable because connecting the light
curve between 0.16 and 0.7 day smoothly produced a flat evolution,
indicating long lasting ($\sim$0.6 days) energy injection.
Using multicolor optical data between 0.6 day and 8 days and the radio
8.5 GHz radio data, we performed light curve modeling.
We examined the light curves for various values of $\theta_{jet}$ and
$\theta_{obs}$. The range of $\theta_{jet}$ covered most of the
estimated values for classical GRBs (from 0.04 rad to 0.5 rad), while
$\theta_{obs}$ varied from 0.0 rad to 1.5 rad when an at least 3-fold
enhancement of $\theta_{jet}$ was examined.
Figure \ref{fig:lc} shows the best-fit model functions that describe
the achromatic rebrightening, $R/Rc$-band temporal evolution, and
radio brightness. The derived burst parameters are also presented in
Table \ref{tab:param}. There are two notable features. The first is
the jet opening half angle $\theta_{jet}$ of 0.1 rad, which is
consistent with those of classical hard GRBs \citep[see Figure 6
  in][]{fong12}. The second is the large observing angle
($\theta_{obs}\sim0.21$ rad), which corresponds to
$\theta_{obs}\sim2\theta_{jet}$.

To compare with other samples such as an off-axis origin XRF
candidate, classical hard GRB, XRR, and the on-axis orphan GRB
candidates, we considered 080330, 990510, 120326A, 131030A, and
PTF11agg (Table \ref{tab:param}). The burst parameters for these
bursts, except 080330, were also determined by using the same box fit
code on the basis of multifrequency afterglow observations
\citep{boxfit,120326a,131030a,cenko13}.
To compare the burst parameters of XRF080330 with those of XRF020903
and others, we also performed the multiband light curve modeling with
the same code by using available multiband photometric results
\citet{080330}.
XRF080330 was detected by {\it Swift} and an afterglow in optical and
IR bands showed the achromatic brightening phase that could be
interpreted using the off-axis jet model \citep{080330}.
However, $E^{src}_{peak}$ was not well characterized, as being less
than 88 keV \citep{080330}, because of the relatively high and narrow
energy band of {\it Swift}/BAT. In addition, the large uncertainties in
the of fluence ratio of the prompt emission ($1.5^{+0.7}_{-0.3}$) were
still compatible with an XRR category based on the modified XRF/XRR
definition for {\it Swift}/BAT \citep{sakamoto08}.
As shown in Figure \ref{fig:080330}, the achromatic brightening light
curves were well modeled, and the derived burst parameters are
presented in Table \ref{tab:param}. One of the key features was
$\theta_{obs}$(=0.12 rad), which was equal to $\theta_{jet}$(0.12
rad). Thus, the achromatic brightening is explained by the off-axis jet
model. However, our viewing angle ($\theta_{obs}\sim\theta_{jet}$) was
not larger as estimated by \citet{080330}
($\theta_{obs}\sim1.5-2\times\theta_{jet}$).

This enables a comparison of the burst parameters, which are presented
in Table \ref{tab:param}.  We also list the observed features of the
prompt emission.  The observed values are widely distributed to
represent the variety of GRB classes. By contrast, the parameters
obtained through the afterglow modeling are comparable to each other,
except for $\theta_{obs}$. Thus, the off-axis jet model is suitable
for explaining the diverse afterglow light curves and the GRB
category.
 
\begin{deluxetable*}{ccccccc}
\tablewidth{0pt}
\tabletypesize{\scriptsize} 
\tablecaption{Summary of burst parameters obtained by observations and numerical modeling\label{tab:param}}
\tablehead{
\colhead{Parameters}&
\colhead{020903} &
\colhead{080330} &
\colhead{990510}&
\colhead{131030A} &
\colhead{120326A}&
\colhead{PTF11agg} 
}
\startdata
Category  &  XRF  &  XRF(XRR?)  &  GRB  &  GRB  &  XRR  &  on-axis orphan(?) \\
$E^{src}_{peak}$ (keV)  &  3.3$^{+1.8}_{-1.0}$  &  $<88$  &  423$^{+42}_{-42}$  &  $406\pm22$  &  107.8$^{+15.3}_{-15.3}$  & $-$ \\
$E_{iso}$ (erg)        &  1.4$^{+18.0}_{-0.7}\times10^{49}$  &  $<2.2\times10^{52}$  &  2.1$^{+0.3}_{-0.3}\times10^{53}$  &  $3.0^{+2.0}_{-0.2}\times10^{53}$ &  3.2$^{+0.4}_{-0.3}\times10^{52}$  & $-$ \\
$z$  &  0.251 &  1.51  &  1.619  &  1.293  &  1.798  & 0.5$<z<$3.0 \\
\hline
$\theta_{jet} $ (rad)  &  0.10 &  0.12   &  0.075  &  0.15  &  0.14  &  0.20 \\
$E$ (erg)      &  $5.9\times10^{52}$ &  $2.3\times10^{52}$  &  $1.8\times10^{53}$  &  $3.4\times10^{52}$  &  $3.9\times10^{52}$  &  $9\times10^{52}$ \\
$n$ (cm$^{-3}$)        &  1.1 &  9.0  &  0.03   &  0.3  &  1.0  &  0.001  \\
$\theta_{obs}$ (rad)   &  0.21 &  0.12  &  0 (fixed)  &  0 (fixed)  &  0 (fixed)  &  0.19 \\ 
$p$                   &   2.8 &  2.1  &  2.28  &  2.1  &  2.5 (fixed)  &  3.0  \\
$\epsilon_{B}$        &  $1.4\times10^{-3}$ &  $1.6\times10^{-1}$  &  $4.6\times10^{-3}$  &  $4.4\times10^{-2}$  &  $1.0\times10^{-3}$  &  $4\times10^{-2}$ \\
$\epsilon_{e}$        &  $2.9\times10^{-1}$ &  $1.4\times10^{-1}$  &  $3.7\times10^{-1}$  &  $2.7\times10^{-1}$  &  $6.9\times10^{-1}$  &  $2\times10^{-1}$ \\
{\bf $\chi^{2}/dof$} & 90.9/9 (10.1) & 512.5/125 (4.1) & 1267.2/198 (6.4) & $-$ & 28.0/20 (1.4) & $-$\\
\hline
Data  &  Opt, Radio &  Opt  &  X,Opt,Radio  &  Opt, ALMA  &  Opt  &  Opt, Radio \\
Ref.   & This work &  This work  &  \citet{boxfit}  &  \citet{131030a}   &  \citet{120326a}  &  \citet{cenko13} 
\enddata
\end{deluxetable*}

 
\subsection{Small values of $E^{src}_{peak}$ 
and $E_{iso}$ with the Off-axis model}

Using the values of $\theta_{jet}$ and $\theta_{obs}$ obtained by
fitting afterglow light curves (Table~\ref{tab:param}), 
we discuss on observed small
values of $E^{src}_{peak}$ and $E_{iso}$ to verify the
off-axis jet model.
We adopted a simple model with a top-hat profile of the 
prompt emission of relativistic jet \citep{granot,woods,ioka,offaxis}.
Following the formalism derived by \citet{graziani} and
\citet{donaghy}, the peak energy 
$E^{src}_{peak}(\theta_{obs})$ and the isotropic energy
$E_{iso}(\theta_{obs})$ were analytically derived as functions of 
$\theta_{obs}$, $\theta_{jet}$, and the Lorentz factor of the jet
$\gamma=(1-\beta^2)^{-1/2}$. 
We defined the ratios $R_1$ and $R_2$ as
\begin{eqnarray}
R_1 &=&\frac{E^{src}_{peak}(\theta_{obs})}{E^{src}_{peak}(0)}\nonumber\\
&=& \frac{2(1-\beta)(1-\beta \cos\theta_{jet})}{2-\beta(1+\cos \theta_{jet})}\nonumber\\
&&\times\frac{f(\beta - \cos\theta_{obs})-f(\beta \cos\theta_{jet}-\cos\theta_{obs})}
{g(\beta-\cos\theta_{obs})-g(\beta \cos\theta_{jet}-\cos\theta_{obs})}~~,\\
R_2 &=&\frac{E_{iso}(\theta_{obs})}{E_{iso}(0)} \nonumber\\
&=& \frac{(1-\beta)^2(1-\beta\cos\theta_{jet})^2}
         {\beta(1-\cos\theta_{jet})[2-\beta(1+\cos\theta_{jet})]}\\ \nonumber
&&\times \left[ f\left(\beta - \cos \theta_{obs}\right)
         -f\left(\beta \cos\theta_{jet}-\theta_{obs}\right)\right]~~,
\end{eqnarray}
where 
\begin{eqnarray}
&&  f(z) = \nonumber\\
&&  \frac{\gamma^{2}(2\gamma^{2}-1)z^{3}+(3\gamma ^{2}\sin^2\theta_{obs}-1)z 
               +2\cos\theta_{obs}\sin^{2}\theta_{obs}}
     {|z^{2}+\gamma^{-2}\sin^{2} \theta_{obs}|^{\frac{3}{2}}}~~,\\
&&g(z) = \frac{2\gamma^{2}z + 2\cos\theta_{obs}}
     {|z^{2}+\gamma^{-2}\sin^{2}\theta_{obs}|^{\frac{1}{2}}}~~.
\end{eqnarray}
For fixed $\theta_{jet}$ and $\gamma$, it can be seen that for
$\theta_{obs}>\theta_{jet}$ \citep[see Figure~2 of][]{donaghy},
both
$E^{src}_{peak}$ and $E_{iso}$ decreases with an increase in
$\theta_{obs}$ because of the relativistic beaming effect.

We consider the case of XRF~020903.  In the following, we fix
$\theta_{obs}=0.21$ rad and $\theta_{jet}=0.1$ rad.
Subsequently, ratios $R_1$ and $R_2$ are calculated for given $\gamma$.
For example, we find 
$R_1=3.45\times10^{-3}$ and $R_2=1.23\times10^{-6}$ if $\gamma=100$.
%
Classical hard GRBs typically have $E^{src}_{peak}=500$~keV \citep{nava},
and therefore we first assume $E^{src}_{peak}(0)=500$~keV.
We then find that $E^{src}_{peak}(\theta_{obs}=0.21)=4.2$~keV, which
is close to the observed result for XRF~020903.
Furthermore, for the observed value of $E_{iso}$,
we consider $E_{iso}(\theta_{obs}=0.21)=1.4\times10^{49}$erg,
resulting in $E_{iso}(0)=1.9\times10^{54}$erg.
In summary,
if a jet with $\theta_{jet}=0.1$ and $\gamma=100$ is seen
on-axis ($\theta_{obs}=0$), we would have
$E^{src}_{peak}=500$~keV and $E_{iso}=1.9\times10^{54}$erg,
which is almost consistent with the ${E^{\rm src}_{\rm peak} - E_{\rm iso}}$ relation \citep{amati,nava}.
The empirical relation could be an indicator of GRB category although the background physics are not yet fully understood. In fact, classical hard GRBs show the relation, but short GRBs do not exhibit it.
Hence, the observed small values of $E^{src}_{peak}$ and
$E_{iso}$ for XRF~020903 are naturally explained by the off-axis jet model.  

We also discuss parameter dependence.
In the following, we fix $\theta_{obs}=0.21$ rad, $\theta_{jet}=0.1$ rad,
and $E_{iso}(\theta_{obs})=1.4\times10^{49}$erg.
The observed best fit value $E^{src}_{peak}(\theta_{obs}=0.21)=3.3$~keV
is reproduced for $\gamma=112$ and $E^{src}_{peak}(0)=500$~keV,
and we obtain $E_{iso}(0)=3.0\times10^{54}$erg.
For $E^{src}_{peak}(0)=250$~keV,
we need a smaller $\gamma$ value (79) is required to obtain 
$E^{src}_{peak}(\theta_{obs}=0.21)=3.3$~keV,
and we find $E_{iso}(0)=7.6\times10^{53}$erg.
Similarly, when we consider $E^{src}_{peak}(0)=1$~MeV,
we have $\gamma=160$ and $E_{iso}(0)=1.3\times10^{55}$erg.
For these cases, the on-axis values, $E^{src}_{peak}(0)$ and
$E_{iso}(0)$ remain within the 3$\sigma$ scatter 
of the ${E^{\rm src}_{\rm peak} - E_{\rm iso}}$ relation.
%
%
These values of $\gamma$ are consistent with those
 measured using the onset of optical afterglows. \citet{liang}
 reported $\gamma$ measurements that were distributed from $\sim90$ to
 $\sim600$ as well as a significant proportion of events that
 exhibited values below 200.  However, the values are lower than the
 prediction ($\gamma>300$ for all GRBs) made by \citet{donaghy} for
 reducing the number of unseen events away from the observed ${E^{\rm src}_{\rm
     peak} - E_{\rm iso}}$ and ${E^{src}_{peak} - E_{\gamma}}$
 \citep{g04} relations.  Here, $\rm E_{\gamma}$ is the jet
 collimation-corrected energy.
The simplest off-axis jet models for XRFs \citep{offaxis,yamazaki}
adopt low values of $\gamma$ ($\sim100$) and predict a large number of
events away from these relations that are not observed (see Figures
4-9 in \citet{donaghy}).  Therefore something more complicated
procedures must be required. For one of examples, \citet{donaghy}
described that a inverse correlation between $\gamma$ and the opening
solid angle of the GRB jet has the effect of greatly reducing the
visibility of off-axis events.Another possibility is to consider more
complicated jet structure \citep[e.g.][]{salafia2015}.

  The spectral peak energy of XRF020903 is conservatively determined
  only the upper limit, reported by \citet{020903}.  Although
  XRF020903 also followed the $E^{src}_{peak}-E_{iso}$ relation when
  we employed the value estimated by the constrained Band function
  with the 90\% confident level, four of the five {\it HETE-2} XRFs
  (including XRF 020903) with the lowest upper limits on
  $E^{obs}_{peak}$ were potential outlier events of the
  $E^{src}_{peak}-E_{iso}$ and the $L_{iso}-E^{src}_{peak}$ relations.
  Because these events are outliers in the $E^{obs}_{peak}-F^{p}_{N}$
  (Figure 1 in \citet{sakamoto05} and Figure 9 in \citet{lamb05}) and
  the $E^{obs}_{peak}-S_{E}$ planes \citep[Figure 2 in ][]{donaghy}.
  Here, $F^{p}_{N}$ and $S_{E}$ denote the peak photon number flux of
  the burst and the energy fluence, respectively.  These results
  indicate that the nature of these four XRFs $-$ and therefore the
  nature of XRFs in general $-$ might differ considerably from those
  of the rest of the XRFs, XRRs, and GRBs. However, our result on one
  of the four outlier events contradicts this suggestion. Hence, we
  may be able to explain the remaining three lowest $E^{obs}_{peak}$
  events in an identical manner, although additional adjustments on
  the jet geometry or the initial Lorentz factor could be required.


\subsection{Prospect with optical surveys}

\begin{figure*}
\epsscale{0.6}
\plotone{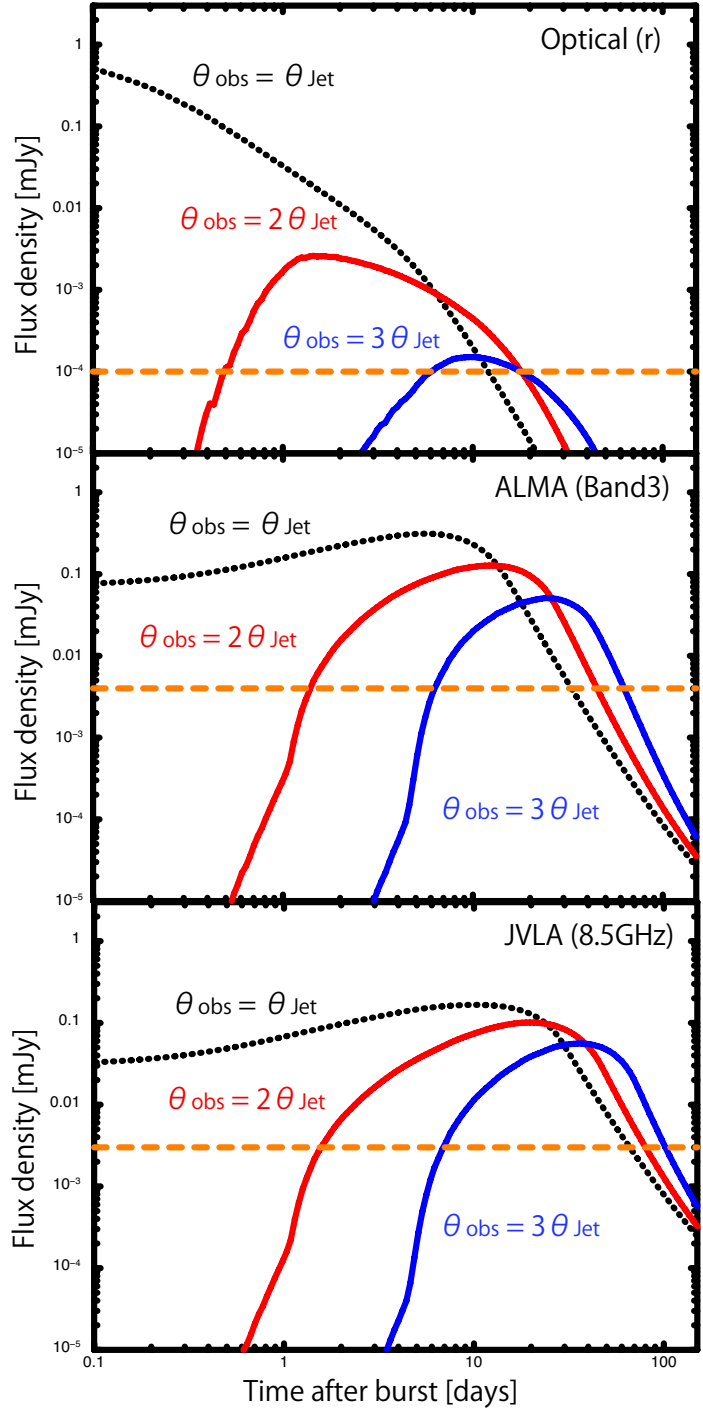}
\caption{Expected off-axis GRB afterglow light curves with observing angle of $2\theta_{jet}$ (red solid line) and $3\theta_{jet}$ (blue solid line) at $z=1$ in the $r'$-band (top), ALMA Band3 (middle), and JVLA 8.5 GHz (bottom) obtained from the off-axis jet modeling of XRF020903. The dash orange lines indicate the sensitivity limit of Subaru ($r'$-band), ALMA (Band3), or JVLA (8.5GHz) with 1 h exposure. }
\label{fig:orphan}
\end{figure*}

Finding off-axis orphan afterglows through untargeted optical surveys
is also a crucial method for establishing a unified picture of GRBs.
Several optical time-domain surveys performed using telescopes with
diameters in the range of 1-2m (e.g. iPTF, Pan-STARRS1) have reported
intriguing new discoveries related to stellar explosions. However, the
sensitivities of these surveys were insufficient for detecting faint
off-axis orphan afterglows. Hence, off-axis orphan GRB afterglows are
yet to be observationally confirmed.  However, a new
wide-field-of-view camera $--$Hyper-Suprime-Cam (HSC)$--$ attached to the
Subaru telescope and the planned LSST have considerable capabilities
to detect the first off-axis orphan GRB afterglow in untargeted
time-domain surveys.

One of the challenges in generic transient surveys is determining
candidates from the various types of optical transients, because the
occurrence of GRB orphan afterglows is rarer compared with that of
known types of supernova. These candidate selection can be performed
using optical photometric survey data and a proper photometric
transient classification. The photometric classification involves
seven steps: (1) finding transient candidates by generating
differential images, (2) generating light curves for transient
components, (3) identifying host galaxies, (4) determining a transient
location in their hosts, (5) matching the candidates with known
sources in various catalogs, (6) matching light curves and color
evolution, and (7) estimating the photometric redshift of hosts.  For
Step (1) and (2), we can employ the algorithm \citep{alard} as
described in \S4 and \citet{sdftran}. For Step (3), (4), and (7),
a considerable number of GRB host galaxy observations have been
performed in the optical and near-infrared range. Systematic unbiased
observations have also been performed using VLT \citep{tough}. The
brightness range of host galaxies for GRBs for $z<2$ is
23.0$--$26.5 mag \citep[e.g.,][]{grbhost}. Since the redshift
range of a considerable number of orphan GRB afterglows with HSC surveys
extends up to $z\sim2$, HSC images (e.g. reference images for PSF matched
subtraction) are sufficiently deep to detect these host galaxies. Thus, we
can also perform the photometric redshift for the host galaxies.
Photometric redshift for GRB host galaxies is also effective
\citep[e.g.,][]{grbphotz}.  Hence, the light curve expectations basis
of off-axis jet origin of XRFs are crucial to establish a effective candidate searches.

Figure \ref{fig:orphan} shows the expected off-axis afterglow light
curves at $z=1$ along with various observing angles ($\theta_{obs}=
\theta_{jet}$, $2\theta_{jet}$, and $3\theta_{jet}$) in the optical
$r$-band, ALMA Band3, and JVLA 8.5 GHz, obtained from the light curve
modeling of XRF020903.  The orange dashed lines in Figure
\ref{fig:orphan} indicate the sensitivity limit for 1 h exposure for
each instrument. Hence, off-axis orphan GRB afterglows (up to
$\sim3\theta_{jet}$) can be detected in optical time-domain surveys by
using 8m class telescopes. Follow-up radio observations are also
crucial for confirmation of orphan afterglows and identification of
constraints on their physical parameters, as shown in Figure
\ref{fig:orphan}.  Because radio temporal evolution is substantially
slower than that of optical temporal evolution, long-term monitoring
by using ALMA and/or JVLA with reasonable exposure ($\sim$1h) requires
the confirmation of the optical candidates.

\section{Conclusion}

We studied XRFs on the basis of redshift measurements, multifrequency
afterglow modeling, and spectral peak measurements ($E^{src}_{peak}$)
to verify the off-axis jet model and to provide feedback to ongoing and
planned optical time-domain surveys, which have considerable potential
for detecting off-axis orphan GRB afterglows. Because off-axis orphan
GRB afterglows are produced as a natural consequence of GRB jets
production, the confirmation of the off-axis origin of XRFs is the
necessary to conduct off-axis orphan GRB afterglow surveys.

To verify the off-axis jet model, we selected XRF020903 by considering the
three aforementioned sample selection factors. For this event, we
reduced the archived data of Subaru to describe the optical light
curves, and found achromatic rebrightening at $\sim$0.7 days. Using
these optical results and radio data obtained from literature, we
performed afterglow light curve modeling with the boxfit code and
found that the off-axis jet model ($\theta_{obs}\sim2\theta_{jet}$) could
explain the achromatic rebrightening, $R/Rc$-band temporal evolution,
and radio brightness.

We also compared the burst parameters of XRF020903 with those of other
categories of events, such as a classical hard GRB, an XRR, and an
on-axis orphan GRB candidate.  For XRF080330, we performed light curve
modeling in a manner similar to that used for XRF020903 by using
optical data from literature, and we confirmed that the off-axis jet model
($\theta_{obs}\sim\theta_{jet}$) could describe the optical afterglow
light curves. We also listed the observed features of the prompt
emission for each event.  The observed values were too widely distributed
to represent the classical hard GRB, XRR, and XRF. By contrast, the
parameters obtained from the afterglow modeling were comparable to each
other, except for $\theta_{obs}$. Thus, the
off-axis jet model was found to be suitable for explaining the diverse afterglow
light curves and the GRB category.

We also verified the observed small values of $E^{src}_{peak}$ and
$E_{iso}$ by adopting a simple model with a top-hat profile of the
prompt emission of a relativistic jet.  The parameters
$E^{src}_{peak}(\theta_{obs})$ and $E_{iso}(\theta_{obs})$ were
analytically derived as functions of $\theta_{obs}$, $\theta_{jet}$,
and $\gamma=(1-\beta^2)^{-1/2}$.  By fixing $\theta_{obs}$ and
$\theta_{jet}$ as 0.21 and 0.1 rad, respectively, we evaluated
$\gamma$, $E^{src}_{peak}$, and $E_{iso}$ observed from the on-axis of
the jet ($E^{src}_{peak}(0)$ and $E_{iso}(0)$). These expected values
were consistent with those of classical hard GRBs, and the observed
small values of $E^{src}_{peak}$ and $E_{iso}$ of XRF~020903 could be
naturally explained by the off-axis jet model.

Finally, we expected off-axis orphan GRB afterglow light curves at
$z=1$ along with three viewing angles on the basis of the XRF
afterglow light curve modeling. To detect these light curves,
especially afterglows with a larger viewing angle
($\theta_{obs}>2\theta_{jet}$), an 8-m class telescope with wide-field
imagers, such as the LSST and Subaru/HSC, is required. Off-axis orphan
GRB afterglows up to $\sim3\theta_{jet}$ can be discovered by
performing time-domain surveys with an 8-m class telescope. Because
such optical time-domain surveys also detect numerous other optical
transients, we presented expected radio afterglow light curves for the
confirmation and determination of burst parameters. Radio light curves
can be monitored using ALMA and JVLA with reasonable exposure.


\acknowledgments
This work is partly supported by the Ministry of Science and
Technology of Taiwan grants NSC 100-2112-M-008-007-MY3 and MOST
103-2112-M-008-021-(YU).
Subaru Suprime-Cam data were acquired through SMOKA, which is operated
by the Astronomy Data Center, National Astronomical Observatory of
Japan.
The PS1 Surveys have been made possible through contributions
of the Institute for Astronomy, the University of Hawaii, the
Pan-STARRS Project Office, the Max-Planck Society and its
participating institutes, the Max Planck Institute for Astronomy,
Heidelberg and the Max Planck Institute for Extraterrestrial Physics,
Garching, The Johns Hopkins University, Durham University, the
University of Edinburgh, Queen's University Belfast, the
Harvard-Smithsonian Center for Astrophysics, and the Las Cumbres
Observatory Global Telescope Network, Incorporated, the National
Central University of Taiwan, and the National Aeronautics and Space
Administration under Grant No. NNX08AR22G issued through the Planetary
Science Division of the NASA Science Mission Directorate.


\end{document}